\documentclass[twocolumn,prb]{revtex4-1}
\usepackage{amsfonts}
\usepackage{amsmath}
\usepackage{amssymb}
\usepackage[final,dvips]{graphicx}

\begin{document}

\title{Thermal conductivity in a mixed state of a superconductor at low
magnetic fields}
\author{A. A. Golubov$^{1}$ and A. E. Koshelev$^{2}$}
\affiliation{$^{1}$ Faculty of Science and Technology and MESA+
Institute of Nanotechnology, University of Twente, 7500 AE, Enschede,
The Netherlands }
\affiliation{$^{2}$ Materials Science Division,
Argonne National Laboratory, Argonne, Illinois 60439, USA}

\pacs{PACS number}
\date{\today}

\begin{abstract}
We evaluate accurate low-field/low-temperature asymptotics of the thermal
conductivity perpendicular to magnetic field for one-band and two-band
s-wave superconductors using Keldysh-Usadel formalism. We show that heat
transport in this regime is limited by tunneling of quasiparticles between
adjacent vortices across a number of local points and therefore widely-used
approximation of averaging over the circular unit cell is not valid. In the
single-band case, we obtain parameter-free analytical solution which
provides theoretical lower limit for heat transport in the mixed state. In
the two-band case, we show that the heat transport is controlled by the ratio of
gaps and diffusion constants in different bands. Presence of a weaker second
band strongly enhances the thermal conductivity at low fields.
\end{abstract}

\maketitle

\section{Introduction}

The thermal conductivity of a metal in the superconducting state is markedly
different from its value in the normal state. The physical reason is that
the Cooper pairs give no contribution to the transport of heat. Therefore
heat transport occurs solely due to the quasiparticle excitations controlled
by the energy gap. Due to this reason, measurements of the thermal
conductivity were extensively used during last decades as a tool to probe
the energy gap symmetry in various new superconducting materials.

Application of a magnetic field $H$ provides a way to generate
quasiparticles in a type-II superconductor. In a s-wave
superconductor at $T \ll T_{c}$, the quasiparticles are localized
near the vortex cores and thermal conduction perpendicular to the
magnetic field is due to tunneling between adjacent vortices.
Thermal conductivity in the mixed state of superconductors was
extensively studied in the past experimentally \cite
{LowellSousaJLTP70,VinenPhysica71,KesJLTP75} and theoretically \cite
{PeschZPhysik74,ImaiWattsJLTP77}. Early theoretical work \cite
{PeschZPhysik74,ImaiWattsJLTP77} addressed superconductors in
diffusive regime (dirty limit) in the mixed state with a temperature
gradient applied transverse to the magnetic flux. The electronic
thermal conductivity was calculated using the quasiclassical
time-dependent superconductivity theory, assuming homogeneous
temperature gradient, using the circular-cell approximation, and
averaging over the unit cell. This approach was later extended to
study heat transport in various unconventional superconductors.
Theory for d-wave superconductors was developed in Ref.
\onlinecite{GrafPRB96} for zero magnetic field and in Refs.
\onlinecite{KubertHPRL98,VorontsovVekhterPRB07} for the mixed state.
More recently, with the discovery of magnesium diboride (MgB$_{2}$),
theory was extended to the case of multiband superconductivity
\cite{KusunosePRB02}. Discovery of pnictides, with possibly
unconventional pairing mechanism due to spin fluctuations leading to
the so-called $s_{\pm}$ pairing state, motivated extension of heat
transport theory \cite{MishraPRB09} taking into account multiband
superconductivity and resonant interband impurity scattering. Many
recent experimental studies reconsider old superconductors, such as
NbSe$_{2}$ \cite{BoakninPRL03}, and addressed novel superconductors,
such as borocarbides (LuNi$_2$B$_2$C) \cite{BoakninPRL01},
Sr$_{2}$RuO$_{4}$ \cite{ShakeripourNJP09}, $C_{6}Yb$
\cite{SutherlandPRL07}, heavy-fermion compounds CeIrIn$_{5}$,
CeCoIn$_{5}$ and UPt$_{3}$ \cite{ShakeripourNJP09}, MgB$_{2}$
\cite{SologubenkoPRB02}, pnictides \cite{Tanatar,Reid} and
iron-silicides \cite{Machida}. Being very sensitive to the gap
structure, thermal transport at low magnetic fields varies
considerably for different compounds.

In this paper we reconsider more accurately the problem of the
thermal conductivity across the magnetic fields for s-wave
superconductors. The case of s-wave superconductor is a standard
reference, all other situations are compared with this case.
Surprisingly, an accurate result for the low-field asymptotics of
the thermal conductivity was never derived. The widely-used circular
unit cell approximation does not give correct result in low magnetic
fields because it misses essential physics. Namely, as the local
thermal conductivity is strongly inhomogeneous, the heat transport
is limited by tunneling between adjacent vortices across certain
local points in the vortex lattice unit cell (bottlenecks).
This leads to general low-field asymptotics of the electronic
thermal conductivity, $\kappa \propto \exp(-\beta \sqrt{B_{c2}/B})$,
where $B_{c2}$ is the upper critical field.\cite{VinenPhysica71}
Surprisingly, the theoretical value of the numerical constant
$\beta$ is not available neither for clean nor for dirty s-wave
superconductors.
For clean case, we provide estimate for this numerical constant
using asymptotics of the Bogolyubov wave functions of the localized
states at zero energy and microscopic value of the upper critical
field.
In the dirty case we were able to perform more quantitative analysis using the
Keldysh-Usadel formalism.  We calculate the thermal conductance at low
temperature and low magnetic field for single- and two-band superconductors
in the dirty limit. In this regime we obtain parameter-free analytical
solution in a single-band case which provides theoretical lower limit for
heat transport in the mixed state.
We find that in dirty case the low-field thermal conductivity is
drastically suppressed in comparison with clean case. Further, we
generalize the developed formalism to a two-band superconductor,
taking MgB$_{2}$ as an example.

\section{Tunneling of quasiparticles between vortex cores in clean isotropic superconductor }

The electronic transverse thermal conductivity in mixed state at low
temperatures and fields is determined by the probability quasiparticle
tunneling between the cores of neighboring vortices. In this section we
evaluate this quantity with exponential accuracy for clean isotropic
superconductors. Even though this estimate is very straightforward and could
be done long time ago, to our great surprise, we did not succeed to find it in
the literature.

The Bogolyubov wave function of the localized state in the
vortex core at $E=0$ decays as \cite{WaveFun}
\begin{equation}
\Psi(r)\propto\exp(-r/\xi_{\Delta}),\text{ }\xi_{\Delta}=v_{F}/\Delta
\label{WaveFunDecay}
\end{equation}
where $\Delta$ is the superconducting gap and $v_{F}$ is the Fermi velocity.
Therefore, the probability of tunneling between the vortex cores separated by
distance $a=\sqrt{2\Phi_{0}/\sqrt{3}B}$ can be estimated as
\begin{equation}
P\!\propto\!|\Psi(a)|^{2}\!\propto\!\exp(-2a/\xi_{\Delta})\!=\!\exp\left(  \!-\sqrt
{\frac{8\Phi_{0}}{\sqrt{3}B\xi_{\Delta}^{2}}}\right)  \label{TunProb}
\end{equation}
The upper critical field $B_{c2}$ for a clean isotropic superconductor at $T=0$
is given by \cite{Hc2Clean}
\begin{equation}
B_{c2}=\frac{e^{2}}{4\gamma}\frac{\Phi_{0}}{2\pi\xi_{0}^{2}},\ \xi_{0}
=\frac{v_{F}}{2\pi T_{c}}\label{Bc2clean}
\end{equation}
with $\gamma=\exp(0.5772)=1.781$ being the Euler constant. Using also the BCS
relation $\pi T_{c}=\gamma\Delta$, we evaluate
\[
2\sqrt{\frac{2\Phi_{0}}{\sqrt{3}B_{c2}\xi_{\Delta}^{2}}}=\sqrt{\frac{16\pi
}{\sqrt{3}\gamma e^{2}}}\approx1.486.
\]
This allows us to represent the tunneling probability (\ref{TunProb}) as
\begin{equation}
P\!\propto\!\exp\left( \! -\sqrt{\frac{16\pi}{\sqrt{3}\gamma e^{2}}\frac{B_{c2}}{B}
}\right)\!  \approx\!\exp\left(  \!-1.486\sqrt{\frac{B_{c2}}{B}}\right).
\label{TunProbClean}
\end{equation}
This result also determines the low-field asymptotics of the
electronic thermal conductivity of clean isotropic superconductor.
The constant in the exponent is obviously sensitive to anisotropy of
Fermi surface. A more quantitative analysis which would include also
evaluation of the preexponential factor requires much more
complicated microscopic kinetic theory for clean limit. In the
following sections we make quantitative calculations of thermal
transport at low fields for dirty superconductor.

\section{The formalism: thermal transport in dirty superconductors within
quasiclassical Keldysh-Usadel model}

Our study is based on the quasiclassical Keldysh-Usadel formalism
\cite{LO,Rammer,Belzig} which was developed to describe
nonequilibrium properties of dirty superconductors. Below we will
reproduce the main relations of this formalism needed for our
derivations. Within this formalism a superconductor is described by
the Green's function
\begin{equation}
G=\left(
\begin{array}{cc}
{\hat{G}}^{R} & {\hat{G}}^{K} \\
0 & {\hat{G}}^{A}
\end{array}
\right) ,
\end{equation}
where in dirty limit the retarded (advanced) Green's functions $G^{R(A)}$
satisfy the Usadel equation\cite{Usadel}
\begin{equation}
-\hbar D\nabla ({\hat{G}}^{R(A)}\nabla {\hat{G}}^{R(A)})=\left[ iE\hat{\tau}
_{3}+\hat{\Delta},{\hat{G}}^{R(A)}\right] .
\end{equation}
Here $E$ is quasiparticle energy, $D$ is the electronic diffusion
coefficient, $\Delta $ is the pair potential
\begin{equation}
\hat{\Delta}=\left(
\begin{array}{cc}
0 & \Delta \\
\Delta ^{\ast } & 0
\end{array}
\right) ,
\end{equation}
${\hat{G}^{K}}={\hat{G}^{R}}\hat{f}-\hat{f}{\hat{G}}^{A}$, $\hat{f}$ is the
distribution function, ${\hat{G}^{R}=\hat{\tau}_{3}G+\hat{\tau}_{1}F}$, ${
\hat{G}^{A}=-\hat{\tau}_{3}\hat{G}^{R\dag }\hat{\tau}_{3}}$, where $G$ and $
F $ are the normal and anomalous Green's functions and ${\hat{\tau}_{i}}$
are Pauli matrices.

Thermal current is given by
\begin{equation}
\mathbf{J}_{th}=\frac{N_{0}D}{{4}}\int{E\mathrm{Tr}}\left[ {\hat{\tau}
_{3}\left( {\hat{G}^{R}\nabla\hat{G}^{K}+\hat{G}^{K}\nabla\hat{G}^{A} }
\right) }\right] {dE},  \label{current}
\end{equation}
where $N_{0}$ is the normal density of states the Fermi level. Writing $\hat{
f}=f_{L}\hat{1}+f_{T}\hat{\tau}_{3}$ , where ${f_{L}}$ and $f_{T}$ are odd
and even in energy components of the distribution function, one can rewrite
the thermal current in the form \cite{Chandra}
\begin{equation}
\mathbf{J}_{th}=N_{0}\int{E}\left[D_{L}\left(E\right) \nabla
f_{L}\left(E\right) +\mathrm{Im}\mathbf{J}_{E}f_{T}\right] {dE} ,  \label{IL}
\end{equation}
where
\begin{align*}
D_{T}&=D\left[\left(\mathrm{Re}G\right)^{2}+\left( \mathrm{Re} F\right) ^{2}
\right], \\
D_{L}&=D\left[ \left( \mathrm{Re}G\right) ^{2}-\left( \mathrm{Im}F\right)
^{2}\right]
\end{align*}
are the energy-dependent spectral diffusion coefficients and $\mathrm{Im}
\mathbf{J}_{E}$ is the spectral supercurrent given by $\mathrm{Im}\mathbf{J}
_{E}=\frac{1} {4}{\mathrm{Tr}}[{\hat{\tau}_{3}({\hat{G}^{R}\mathbf{\nabla}
\hat{G}^{R} -\hat{G}^{A}\mathbf{\nabla}\hat{G}^{A}})}]=\mathrm{Im}F^{R}
\mathrm{Re} F^{R}{\nabla\chi}$, where ${\chi}$ is the superconducting phase.
In the mixed state all quantities are spatially inhomogeneous (coordinate
dependences are dropped for brevity). Functions $f_{L}$ and $f_{T}$ satisfy
the following kinetic equations \cite{Rammer,Belzig}
\begin{align}
\nabla\left( D_{T}\nabla f_{T}\right) +\mathrm{Im}\mathbf{J}_{E}\nabla f_{L}
& =2Rf_{T},  \label{fT} \\
\nabla\left( D_{L}\nabla f_{L}\right) +\mathrm{Im}\mathbf{J}_{E}\nabla f_{T}
& =0.  \label{fL}
\end{align}
where $R=\frac{1}{4}Tr\left[ \hat{\Delta}({\hat{G}^{R}+\hat{G}^{A}})\right] $. In thermal equilibrium $f_{L}=\tanh\left( E/2k_{B}T\right)$ and $f_{T}=0$.
The thermal conductivity components are defined as
\begin{equation}
\kappa_{\alpha}=\langle J_{th,\alpha}\rangle/\langle\nabla_\alpha T \rangle,
\end{equation}
where $\langle{\mathbf{\nabla}} T\rangle$ is the average temperature
gradient and $\alpha=(x,y,z)$.

In the following, we shall use the standard $\theta$-parametrization, ${\hat{
G}^{R}}(\mathbf{r})=\hat{\tau}_{3}\cos[\theta(\mathbf{r})]+\hat{\tau}
_{1}\sin [\theta(\mathbf{r})]$ in which the Usadel equation has the
following form \cite{Watts,GK}
\begin{equation}
D\left(\mathbf{\nabla}^{2}\theta-p^{2}\cos\theta\sin\theta\right)+2\Delta
\cos\theta +2iE\sin\theta=0  \label{Usadel}
\end{equation}
where $\mathbf{p}(\mathbf{r})=\mathbf{\nabla}\phi-(2\pi/\Phi_{0})\mathbf{A}$
is superconducting momentum (within circular cell approximation $
p=1/r-r/r_{s} ^{2}$, $r_{s}=\sqrt{\Phi_{0}/\pi H}$). The selfconsistency
equation has the form
\begin{equation}
\Delta\ln\frac{T}{T_{c}}=\!\frac{T}{T_{c}}\sum_{\omega}\!\left( \sin \theta-
\frac{\Delta}{\omega}\right) \!  \label{Usadel_selfconsist}
\end{equation}
The diffusion coefficients in the $\theta$-parametrization are given by the expressions
$D_{T}(E,\mathbf{r})=D\cosh ^{2}\left(\mathrm{Im}[\theta(E,\mathbf{r})]\right)$, $
D_{L}(E,\mathbf{r})=D\cos^{2}\left(\mathrm{Re}[\theta(E,\mathbf{r})]\right)$.

\section{Thermal transport in the vortex state at low fields}

\begin{figure}[ptb]
\begin{center}
\includegraphics[width=3.5in]{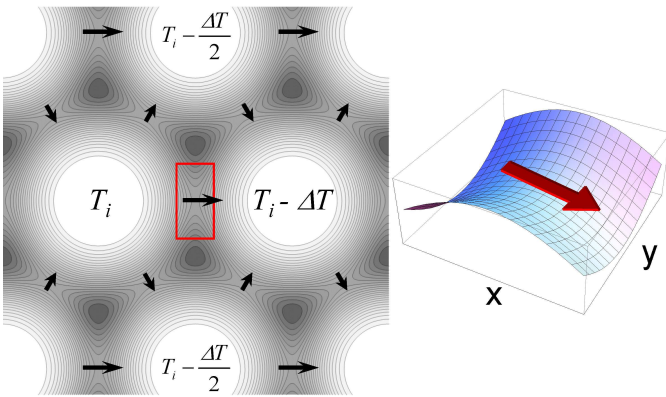}
\end{center}
\caption{\emph{Left:} Graylevel map of the local thermal resistance $\propto
1/D_{L}(0,\mathbf{r})$ in the vortex lattice. Light regions correspond to
low thermal resistance. Arrows illustrate heat flow in the bottleneck
regions. \emph{Right:} Three-dimensional plot of the local thermal
resistance in the bottleneck region marked by rectangle in the left picture.}
\label{Fig-LattBottlenecks}
\end{figure}

We consider a superconductor in low magnetic field, $B\ll B_{c2}$. We assume
an ideal triangular vortex lattice of vortex lines and study thermal
transport in the direction perpendicular to the field. The local thermal
conductivity is mostly determined by the diffusion constant $D_{L}(E,\mathbf{
\ r})$ at low energies. This quantity is very inhomogeneous in the vortex
state. It has maxima at the vortex cores and rapidly drops away from the
cores reflecting localization of quasiparticles in the core regions. This
means that the local thermal resistance $\propto 1/D_{L}(0,\mathbf{r})$ is
maximal at the boundaries of the lattice unit cell. In such situation
the temperature is mostly homogeneous within the unit cells and only changes
in the boundary regions between the cells, see Fig. \ref{Fig-LattBottlenecks}. This means that thermal transport occurs via ``bottlenecks``, saddle points
of $D_{L}(E,\mathbf{r})$. Our purpose is to evaluate average thermal
conductivity limited by these bottlenecks.
%At low magnetic field thermal transport is determined by energy flow via
%bottlenecks, saddlepoints of $D_{L}(E,\mathbf{r})$ at boundaries of the
%lattice unit cell, see Figure \ref{Fig-LattBottlenecks}.
Since in the vicinity of bottlenecks, the spectral supercurrent $\mathrm{Im}
\mathbf{J}_{E}$ vanishes by symmetry, the expression for local energy
current simplifies and has the form: $\mathbf{J}_{th}(\mathbf{r})\approx
N_{0}\int dEED_{L}\mathbf{\nabla }f_{L}$.

We start with evaluation of the diffusion constant
$D_{L}(E,\mathbf{r})$ which is determined by the real part of the
Green's function $\theta (E, \mathbf{r})$. For an isolated vortex,
at distances $r\!\gg \!\xi $ from its core we can present
$\Delta$ and $\theta$ as $\theta (\mathbf{r})=\theta _{0}+
\tilde{\theta}(\mathbf{r})$, $\Delta
(\mathbf{r})=\Delta_0+\tilde{\Delta}(\mathbf{r})$ where $\Delta_0$
and $\theta _{0}$ are the equilibrium values at zero magnetic field,
$\tan \theta _{0}\!=\!i\Delta _{0}/E$ and $\tilde{\theta
}(\mathbf{r})\!=\!\tilde{\theta}_{r}(\mathbf{r})\!+\!i\tilde{\theta}_{i}
(\mathbf{r})$ is small correction. For $E\ll \Delta _{0}$ , $\theta
_{0}\approx \pi /2\!+\!iE/\Delta _{0}$, therefore the
energy-diffusion constant in this region
$D_{L}(E,\mathbf{r})\!\approx \!D\tilde{\theta} _{r}^{2}$. The real
and imaginary parts of $\tilde{\theta}$ obey the
following equations %\begin{widetext}
\begin{align}
& D\mathbf{\nabla }^{2}\tilde{\theta}_{r}-D\frac{E^{2}+\Delta _{0}^{2}}{
E^{2}-\Delta _{0}^{2}}p^{2}\tilde{\theta}_{r}-2\sqrt{\Delta _{0}^{2}-E^{2}}
\tilde{\theta}_{r}=0,  \label{Eq_thetar} \\
& D\mathbf{\nabla }^{2}\tilde{\theta}_{i}-D\frac{E^{2}+\Delta _{0}^{2}}{
E^{2}-\Delta _{0}^{2}}p^{2}\tilde{\theta}_{i}-2\sqrt{\Delta _{0}^{2}-E^{2}}
\tilde{\theta}_{i}  \notag \\
& =\frac{2E\tilde{\Delta}}{\sqrt{\Delta
_{0}^{2}-E^{2}}}-D\frac{E\Delta _{0}}{ \Delta _{0}^{2}-E^{2}}p^{2},
\label{Eq_thetai}
\end{align}
%\end{widetext}
where $p=1/r$ is the gauge-invariant phase gradient. For an isolated
vortex $\tilde{\theta}_{r}$ and $\tilde{\theta}_{i}$ have
qualitatively different behavior at large distances for $E<\Delta
_{0}$: $ \tilde{\theta}_{i}$ decays as $p^{2}\propto 1/r^{2}$ and
$\tilde{\theta}_{r}$ decays exponentially
\begin{align}
\tilde{\theta}_{v,r}(r)& =C_{v}\frac{\exp (-k_{\xi }r)}{\sqrt{k_{\xi }r}},
\label{AsymptSingleVort} \\
k_{\xi }(E)& =\sqrt{2}\left( \Delta _{0}^{2}-E^{2}\right) ^{1/4}/\sqrt{D}.
\notag
\end{align}
Evaluation of numerical constant requires solution of full nonlinear
problem. Numerical solution by the method described \cite{GK} provides $
C_{v}\approx 4.2$. In the vortex lattice away from the core regions $\tilde{
\theta}_{r}$ can be represented as a sum of contributions from individual
vortices
\begin{equation}
\tilde{\theta}_{\mathrm{lat},r}(\mathbf{r})\!\approx \!\sum_{\mathbf{R}}
\tilde{\theta}_{v,r}\left( \mathbf{r}\!-\!\mathbf{R}\right) \!=\!\sum_{
\mathbf{R}}C_{v}\frac{\exp (-k_{\xi }|\mathbf{r}\!-\!\mathbf{R}|)}{\sqrt{
k_{\xi }|\mathbf{r}-\mathbf{R}|}},  \label{AsymptPhaseLatt}
\end{equation}
where $\mathbf{R}$ are the vortex coordinates. In particular, for triangular
lattice $\mathbf{R}=(a/2+ma+na/2,n\sqrt{3}/2)$ where $a=\sqrt{2\Phi _{0}/
\sqrt{3}B}$ is the lattice constant and $m$ and $n$ are integers.

At small field thermal transport is determined by energy flow via
bottlenecks, saddlepoints of $D_{L}(E,\mathbf{r})$ at boundaries of the
lattice unit cell. Near the bottleneck point, $(x,y)=(0,0)$, we can keep
only contribution from two neighboring vortices located at $(\pm a/2,0)$
which gives
\begin{equation}
\tilde{\theta}_{\mathrm{lat},r}(\mathbf{r})\approx\frac{2C_{v}}{\sqrt{k_{\xi
}a/2}}\exp\left( -k_{\xi}a/2\right) \cosh\left( k_{\xi}x\right)
\exp(-k_{\xi}y^{2}/a)  \label{PhaseBneck}
\end{equation}
and
\begin{equation}
D_{L}(E,\mathbf{r})\approx D\frac{8C_{v}^{2}}{k_{\xi}a}\exp\left( -k_{\xi
}a\right) \cosh^{2}\left( k_{\xi}x\right) \exp(-2k_{\xi}y^{2}/a).
\label{DiffBneck}
\end{equation}
Using also quasiequilibrium approximation for the gradient of $f_L(E,
\mathbf{r})$, $\nabla_x f_L(E, \mathbf{r})\approx
-\cosh^{-2}(E/2k_BT)(E/2k_BT^2)\nabla_x T(\mathbf{r})$, we obtain the heat
flow near the bottleneck $J_{\mathrm{th},x}(\mathbf{r})\approx -\kappa_x
(\mathbf{r})\nabla_{x}T(\mathbf{r})$, where the local thermal conductivity $
\kappa_x(\mathbf{r})$ is given by
\begin{widetext}
\[
%J_{th,x}(\mathbf{r})
\kappa_x(\mathbf{r})\approx 8C_{v}^{2}DN_{0}\int_{-\infty}^{\infty}
dEE\frac{\exp\left(  -k_{\xi}a\right)  }{k_{\xi}a}\cosh^{2}
\left(k_{\xi }x\right)
\exp\left(  -\frac{2k_{\xi}y^{2}}{a}\right)
\frac{E/2k_{B}T^{2} }{\cosh^{2}\left(  E/2k_{B}T\right)}.
%\nabla_{x}T.
\]
\end{widetext}
At low temperatures $T<\Delta_{0}\xi/a$ the main contribution to the energy
integral comes from the region $E\lesssim T$. This allows us to neglect the
energy dependence of $k_{\xi}(E)$ and replace $k_{\xi}(E)\rightarrow k_{\xi
0}=\sqrt{2\Delta_{0}/D}=1/\xi_{\Delta}$. In this case, using $\int_{-\infty
}^{\infty}x^{2}\cosh^{-2}\!x\ dx=\pi^{2}/6$, we obtain
\begin{align}
\kappa_{x}(\mathbf{r}) =\frac{16\pi^{2}}{3}C_{v}^{2}k_{B}^{2}TN_{0}D\frac{
\exp\left( -k_{\xi 0}a\right) }{k_{\xi0}a} \nonumber\\
\ \times\cosh^{2}\left( k_{\xi0}x\right) \exp\left( -\frac{2k_{\xi0}y^{2}}{a}
\right).
\end{align}
%\end{widetext}
Near the bottleneck region the local thermal conductivity has a form $
\kappa(x,y)=\kappa_{0}F_{x}(x)F_{y}(y)$, where the function $
F_{x}(x)=\cosh^{2}\left( k_{\xi0}x\right) $ has minimum at $x=0$ and $
F_{y}(y)=\exp\left( -2k_{\xi0}y^{2}/a\right) $ has maximum at $y=0$. The
total flow per unit length along the field through the bottleneck is given
by
\begin{equation}
I_{th}\approx\kappa_{0}F_{x}(x)\left[ \int_{-\infty}^{\infty}F_{y} (y)dy
\right] \nabla_{x}T.
\end{equation}
Energy conservation requires that $I_{th}$ has to be $x$-independent.
Therefore, the temperature drop $\Delta T$ across the bottleneck can be
evaluated as
\begin{equation}
\Delta T\approx\frac{\int_{-\infty}^{\infty}F_{x}^{-1}(x)dx}{\kappa_{0}
\int_{-\infty}^{\infty}F_{y}(y)dy}I_{th}
\end{equation}
meaning that the total thermal conductance through this region $K=I_{th}
/\Delta T$ can be evaluated as
\begin{equation}
K=\kappa_{0}\frac{\int_{-\infty}^{\infty}F_{y}(y)dy}{\int_{-\infty}^{\infty
}F_{x}^{-1}(x)dx}=\frac{8\pi^{5/2}C_{v}^{2}}{3\sqrt{2}}k_{B}^{2}TN_{0} D
\frac{\exp\left( -k_{\xi0}a\right) }{\sqrt{ak_{\xi0}}}.  \label{BneckThCond}
\end{equation}
Evaluating the total average energy flow density
\begin{equation}
J_{th}=\frac{I_{th}+I_{th}/4}{a\sqrt{3}/2}=\frac{5}{2\sqrt{3}}K\frac{\Delta
T }{a},
\end{equation}
we obtain the final result for the low-field/low-temperature limit for the
thermal conductivity in the vortex-lattice state
\begin{align}
\kappa/T & \approx\frac{10\sqrt{2}\pi^{5/2}C_{v}^{2}}{3\sqrt{3}}k_{B}
^{2}N_{0}D\frac{\exp\left( -k_{\xi0}a\right) }{\sqrt{k_{\xi0}a}}  \notag \\
& =\frac{10\sqrt{2}\pi^{5/2}C_{v}^{2}}{3\sqrt{3}}k_{B}^{2}N_{0}D\frac {
\exp\left( -\sqrt{(8\pi/\sqrt{3})B_{c2}/B}\right) }{\left( (8\pi/\sqrt {3}
)B_{c2}/B\right) ^{1/4}}
\end{align}
where we used relations $B_{c2}=\Phi_{0}k_{\xi0}^{2}/4\pi$ and $k_{\xi 0}a=
\sqrt{(8\pi/\sqrt{3})B_{c2}/B}$.

Introducing the thermal conductivity in the normal state $\kappa_{N}=\frac {
\pi^{2}}{3}k_{B}^{2}N_{0}DT,$ we rewrite the final result in the form
%\begin{widetext}
\begin{align}
\frac{\kappa}{\kappa_{N}}&\approx 10 \sqrt{\frac{2\pi}{3}}C_{v}^{2}\frac {
\exp\left( -\sqrt{(8\pi/\sqrt{3})B_{c2}/B}\right) }{\left( (8\pi/\sqrt {3}
)B_{c2}/B\right) ^{1/4}}  \notag \\
&\approx130\left( \frac{B}{B_{c2}}\right) ^{1/4}\exp\left( -3.81\sqrt{\frac{
B_{c2}}{B}}\right).  \label{ThCondResultNumeric}
\end{align}
%\end{widetext}
This parameter-free analytical result provides theoretical lower
limit for the heat transport in the mixed state in an isotropic
dirty s-wave superconductor at low field. The constant $3.81$ in the
exponent is significantly higher than the constant $1.486$ which we
evaluated for the clean case (\ref{TunProbClean}) meaning that the
scattering drastically suppresses the quasiparticle thermal
conductivity at low fields.

\section{Discussion of experiment}

Rather limited set of experimental data is available on electronic
contribution to the thermal conductivity at low temperatures and low
magnetic fields, since phonon contribution should be accurately
subtracted in this regime. To our knowledge, the only experimental
data on electronic thermal conductivity
which clearly demonstrate low-field/low-temperature behavior
expected for an s-wave superconductor
are reported for Nb in Refs.\
\onlinecite{LowellSousaJLTP70,VinenPhysica71} and for V${_3}$Si in
Ref.\ \onlinecite{BoakninPRL03}. In these experiments the samples
were in the clean limit. In Ref. \onlinecite{VinenPhysica71} the
electronic thermal conductivity was fitted by the expression $\kappa
\propto \exp(-\beta \sqrt{B_{c2}/B})$ for values of $B$ up to about
$B_{c2}/3$, with $\beta=1.66$. This number is slightly higher than
our estimate of $\beta=1.486$ in Eq. (\ref{TunProbClean}). This
difference can be explained by the influence of impurity scattering.
The shape of field dependence of thermal conductivity for V${_3}$Si
reported in Ref. \onlinecite{BoakninPRL03} is in qualitative
agreement with the predicted exponential dependence, however the
quantitative analysis was not made and the value of $\beta$  was not
explicitly extracted.
%
%The data were found in qualitative agreement with the calculations
%assuming the clean limit \cite{Dukan}.
In a field $B=B_{c2}/20$ the
value of $\kappa \approx 2.5\cdot 10^{-3}\kappa_N $ provided in
Ref.\ \onlinecite{BoakninPRL03} exceeds our dirty-limit estimate,
Eq. (\ref{ThCondResultNumeric}), by about two orders of magnitude.
%, while overall field dependence  is
%in qualitative agreement with Eq. (\ref{ThCondResultNumeric}).
This discrepancy can be naturally attributed to the fact that
measured V${_3}$Si samples were in the clean limit.
The magnitude of thermal conductivity at not very small magnetic
fields is in good qualitative agreement with calculations made for
the clean limit using the Landau-level expansion and assuming
homogeneous temperature gradient \cite{Dukan}.
%%Impurity scattering strongly suppresses thermal transport at low
%temperatures.
To our knowledge, there are no data available on electronic thermal
conductivity of dirty s-wave superconductors in the
low-field/low-temperature limit. In available measurements of alloys
\cite{LowellSousaJLTP70} and dirty Nb samples \cite{KesJLTP75} the thermal
conductivity at low fields and temperatures is dominated by phonons
and separating electronic contribution is a challenging task.

\section{Extension to a two-band superconductor}

Here we extend the above formalism to a two-band superconductor. At present,
the most established example of such system is MgB$_{2}$, which is characterized
by two electronic bands: $\pi$-band and $\sigma$-band, see, e.\ g.\, recent
review Ref.~\onlinecite{XiReview08}. The quasi-two-dimensional $\sigma$-band is
characterized by stronger superconductivity than the three-dimensional $\pi$
-band. Heat transport in a two-band superconductor was studied theoretically
in Ref.~\onlinecite{KusunosePRB02} assuming clean limit conditions and using
the averaging over unit cell method \cite{PeschZPhysik74,ImaiWattsJLTP77}.
In the dirty limit, theory of density of states in the mixed state and the
upper critical field for MgB$_{2}$ was developed in Refs.\
\onlinecite{MgB2-DoS,MgB2-Hc2}. Below we extend the calculations of the heat
transport presented above, to the case of a diffusive two-band
superconductor, taking MgB$_2$ as an example.

In the presence of two electronic bands with different energy gaps,
low-temperature behavior of thermal transport is determined by the band with
lower gap ($\pi$-band). Still, the generalization from single-band to
two-band case involves not simply renormalization of the energy gap, but
also correction to the asymptotic behavior of Green's function in the $\pi$
-band and to the upper critical field.

The Usadel equation for the $\pi $-band reads
\begin{equation}
D_{\pi }\left(\mathbf{\nabla }^{2}\theta_{\pi}\!-\!p^{2}\cos
\theta_{\pi} \sin \theta_{\pi}\right )\!+\!2\Delta _{\pi }\cos
\theta_{\pi}\!+\!2iE\sin \theta_{\pi}\!=\!0,
\end{equation}
where $D_{\pi }$ and $\Delta _{\pi }(\mathbf{r})$ are the diffusion constant
and gap for the $\pi$-band. Similar to a single-band case, asymptotics of
the Green's function in the $\pi$-band at large distance from the vortex
core is given by
\begin{equation}
\tilde{\theta}_{\pi ,r}(r)=C_{\pi }\frac{\exp (-k_{\pi }r)}{\sqrt{k_{\pi }r}}
.  \label{piAsympt}
\end{equation}
with
\begin{equation*}
k_{\pi }(E) =\sqrt{2}\left( \Delta _{\pi 0}^{2}-E^{2}\right) ^{1/4}/\sqrt{
D_{\pi }}.
\end{equation*}
The main exponential dependence of the zero-energy Green's function at the
bottleneck point is $\propto \exp (-k_{\pi 0}a)$, where $k_{\pi 0}=k_{\pi
}(0) =\sqrt{2\Delta _{\pi 0}/D_{\pi }}$, which gives
\begin{align}
k_{\pi0}a&=\sqrt{B_{\pi}/B},  \label{kpia} \\
B_{\pi}&=\frac{2\Delta_{\pi0}}{D_{\pi}}\frac{2\Phi_{0}}{\sqrt{3}}.
\label{Bpi}
\end{align}
%\begin{align*}
%k_{\pi }(E)& =\sqrt{2}\left( \Delta _{\pi 0}^{2}-E^{2}\right)
%^{1/4}/\sqrt{
%D_{\pi }}, \\
%k_{\pi 0}& =\sqrt{2\Delta _{\pi 0}/D_{\pi }}.
%\end{align*}
Therefore, the magnetic field dependence of thermal conductivity is
determined by the field scale $B_{\pi}$ and can be presented in the form
similar to Eq.\ (\ref{ThCondResultNumeric}),
\begin{equation}
\frac{\kappa }{\kappa _{\pi N}}\approx 10\sqrt{\frac{2\pi }{3}}C_{\pi }^{2}
\frac{\exp\left(-\sqrt{B_{\pi}/B}\right)}{\left(B_{\pi}/B\right)^{1/4}},
\label{kappaTwoBand}
\end{equation}
where $\kappa _{\pi N}=\frac{\pi ^{2}}{3}k_{B}^{2}N_{\pi }D_{\pi }T$ is the
partial $\pi $-band contribution to the normal-state thermal conductivity.

To proceed further, we have to find relation between the $\pi$-band field
scale $B_{\pi}$ and the upper critical field $B_{c2}$ for a two-band
superconductor. The upper critical field at low temperatures, $T\ll T_{c}$,
is given by \cite{MgB2-Hc2}
\begin{widetext}
%\begin{align}
%B_{c2}(0) &  =a_{c2}B_{c2}^{s}(0),\label{Hc20Exact}\\
%a_{c2} &  =\exp\left(  -\frac{W_{1}+W_{2}-\ln\left(  r_{x}\right)
%}{2} +\sqrt{\frac{\left(  W_{1}+W_{2}-\ln\left(  r_{x}\right)
%\right)  ^{2}} {4}+W_{1}\ln\left(  r_{x}\right)  }\right)\nonumber
%\end{align}
\begin{equation}
B_{c2}(0) =a_{c2}B_{c2}^{s}(0),\ \ a_{c2} =\exp\left(
-\frac{W_{1}\!+\!W_{2}\!-\!\ln r_{x}}{2} +\sqrt{\frac{\left(
W_{1}\!+\!W_{2}\!-\!\ln r_{x} \right) ^{2}} {4}\!+W_{1}\ln r_{x}
}\right), \label{Hc20Exact}
\end{equation}
\end{widetext}
where indices 1 and 2 correspond to the $\sigma$ and $\pi$ bands,
\begin{equation*}
W_{1,2}=\frac{\mp(\Lambda_{11}-\Lambda_{22})/2 +\sqrt{(\Lambda_{11}-
\Lambda_{22})^2/4+\Lambda_{12}\Lambda_{21}}} {\Lambda_{11}\Lambda_{22}-
\Lambda_{12}\Lambda_{21}},
\end{equation*}
$\Lambda_{\alpha \beta}$ is the coupling-constant matrix, $r_{x}=\mathcal{D}
_{\sigma }/\mathcal{D}_{\pi }$, $\mathcal{D}_{\sigma ,\pi }$ are the
diffusion constants in $\sigma $ and $\pi $ bands, %\begin{align}
%B_{c2}^{s}(0)& =B_{\sigma }e^{-\gamma _{E}}/4=\frac{e^{-\gamma _{E}}}{2}
%\frac{T_{c}\Phi _{0}}{\mathcal{D}_{\sigma }}=\frac{\Delta _{BCS}\Phi _{0}}{
%2\pi \mathcal{D}_{\sigma }} \\
%B_{\sigma }& \equiv 2T_{c}\Phi _{0}/\mathcal{D}_{\sigma } \\
%\Delta _{BCS}& =\pi e^{-\gamma _{E}}T_{c}\text{, }e^{-\gamma _{E}}/4\approx
%0.140
%\end{align}
\begin{equation}
B_{c2}^{s}(0)=\frac{\Delta _{BCS}\Phi _{0}}{ 2\pi \mathcal{D}_{\sigma }}
\label{Bc2sigma}
\end{equation}
is the single-band upper critical field for the $\sigma$-band, and $\Delta
_{BCS}=\pi e^{-\gamma _{E}}T_{c}\approx 1.764 T_{c}$ ($\gamma_E\approx
0.5772 $ is the Euler constant).

To make estimates for MgB$_{2}$, we use the following coupling
matrix elements \cite {MgB2-DoS,MgB2-Hc2}:
$\Lambda_{11}\!\approx0.81,\ \Lambda_{22} \!\approx0.278,\
\Lambda_{12}\!\approx0.115,\ \Lambda_{21}\!\approx0.091$, which
gives $\!W_{1}\!\approx0.088$ and $W_{2}\!\approx2.56$. With such
coupling matrix the two-band BCS model gives \cite{MgB2-DoS} $\Delta
_{\pi 0}\simeq 0.3\Delta _{\sigma 0}\simeq $ $0.177\pi T_{c}$. Since
the parameter $W_{1}$ is small, typically the inequality $W_{1}|\ln
r_{x}|\ll \left(W_{2}-\ln r_{x} \right) ^{2}/4$ is valid. In this
case one can expand Eq.\ (\ref{Hc20Exact}) with respect to $W_{1}$
and obtain simple result
\begin{equation}
a_{c2}\approx 1+\frac{W_{1}\ln r_x }{W_{2}-\ln r_x },  \label{HcT0}
\end{equation}
meaning that the upper critical field is close to $B_{c2}^{s}(0)$ and is
mostly determined by the coherence length of the $\sigma$-band. Using Eqs.\
(\ref{Bpi}), (\ref{Hc20Exact}), and (\ref{Bc2sigma}), we obtain the relation
between $B_{\pi}$ and $B_{c2}(0)$
\begin{equation}
B_{\pi}=\frac{8\pi}{\sqrt{3}a_{c2}}\frac {\Delta_{\pi0}}{\Delta_{BCS}}\frac{
\mathcal{D}_{\sigma}}{\mathcal{D}_{\pi} }B_{c2}(0),
\end{equation}
which presents the main result of this section. This scale has to be
compared with the scale $(8\pi/\sqrt{3})B_{c2}(0)$ for the single-band case.
%The main exponential dependence of the Green's function at the
%bottleneck point is $\propto \exp (-k_{\pi 0}a)$, where
%\begin{widetext}
%\[
%k_{\pi0}a=\sqrt{\frac{2\Delta_{\pi0}}{D_{\pi}}\frac{2\Phi_{0}}{\sqrt{3}B}
%}=\sqrt{\frac{4\Delta_{\pi0}}{\sqrt{3}D_{\pi}}\frac{\Phi_{0}}{a_{c2}B_{c2}
%^{s}(0)}\frac{B_{c2}(0)}{B}}=\sqrt{\frac{8\pi}{\sqrt{3}a_{c2}}\frac
%{\Delta_{\pi0}}{\Delta_{BCS}}\frac{\mathcal{D}_{\sigma}}{\mathcal{D}_{\pi}
%}\frac{B_{c2}(0)}{B}}
%\]
%\end{widetext}

In order to determine the pre-exponential factor $C_{\pi }$ in Eqs.\ (\ref
{piAsympt}) and (\ref{kappaTwoBand}), we have to calculate the Green's
function $\tilde{\theta}_{\pi ,r}(r)$, which requires solution of the full
two-band Usadel problem, as described in Ref.\ \onlinecite{MgB2-DoS}. An
important parameter is the ratio of diffusion coefficients in two bands $r_x$
. For illustration, we consider two cases here: $r_x=1$ and $0.2$, for which
the ratios of the coherence lengths in the two bands are $\xi_\pi/\xi_\sigma
=\sqrt{(D_\pi/D_\sigma)(\Delta_{\sigma0}/\Delta_{\pi0})}=1.83$ and 4.1. For
these two cases we compute $B_\pi\approx 0.32 (8\pi/\sqrt{3})B_{c2}(0)$, $
C_{\pi }\approx 3.6$ for $r_x=1$ and $B_\pi\approx 0.065 (8\pi/\sqrt{3}
)B_{c2}(0)$, $C_{\pi }\approx 2.9$ for $r_x=0.2$. This gives
\begin{align}
\ \frac{\kappa }{\kappa _{\pi N}}& \!\approx\! 130\left( \frac{B}{B_{c2}(0)}
\right) ^{1/4}\exp \left( -2.15\sqrt{\frac{B_{c2}(0)}{B}}\right) \ \text{,\ }
r_x=1  \notag \\
& \!\approx\! 130\left( \frac{B}{B_{c2}(0)}\right)^{1/4}\!\exp \left(\! -0.98\sqrt{
 \frac{B_{c2}(0)}{B}}\right) \text{,\ }r_x\!=\!0.2  \notag
\end{align}
We can see that the field scale in the two-band case is strongly reduced in
comparison with the single-band case leading to large enhancement of the
thermal conductivity at low fields. This reduction is mostly caused by the
smaller energy gap in the $\pi$ band. Another factor which may contribute is
possible large value of the diffusion constant $D_{\pi}$. The smaller field
scale caused by the larger coherence length in the $\pi$-band is an
established feature and important fingerprint of the two-band
superconductivity in MgB$_2$. This small scale was experimentally observed
not only in the thermal conductivity \cite{SologubenkoPRB02}, but also in
the specific heat \cite{SpecHeatMgB2} and flux-flow resistivity \cite
{ShibataPRB03}. For the magnetic field applied along c-axis it is 3-5 times
smaller than $B_{c2}$. In this case, the low-field regime described by Eq.\
(\ref{kappaTwoBand}) is expected at fields $<B_{c2}/30\approx 100$G.
Unfortunately, most experimental data of Ref.\ \onlinecite{SologubenkoPRB02}
are presented for higher magnetic fields.

We shall note that in available MgB$_2$ single crystals estimates suggest
that the $\sigma$ band is in the clean limit. However, our results should be
qualitatively applicable even in this case. The reason is that dominant
contribution comes from the $\pi$-band which is in the dirty limit, as
argued in Ref. \onlinecite{MgB2-DoS} where the low-energy DoS in the vortex
state of MgB$_2$ was calculated.

In summary, we have readdressed the problem of the heat transport of
a superconductor in the mixed state at low temperatures and low
magnetic fields, going beyond the circular unit cell approximation.
In the clean limit we estimated the numerical constant $\beta$ in
the low-field asymptotics of the electronic thermal conductivity,
$\kappa \propto \exp(-\beta \sqrt{B_{c2}/B})$, using the Bogolyubov
wave functions of the localized states at zero energy. In the dirty
limit we have performed quantitative analysis of heat transport
using Keldysh-Usadel formalism and have shown that heat transport is
limited by tunneling between adjacent vortices across certain local
points (bottlenecks). In the isotropic s-wave superconductor we have
obtained parameter-free analytical solution which provides
theoretical lower limit for heat transport in the mixed state. Based
on this solution, one can conclude that low-field/low-temperature
thermal conductivity in the mixed state is drastically suppressed by
impurity scattering. We have extended our results to the case of a
two-band superconductor, taking MgB$_{2}$ as an example. In this
case, we predict an enhancement of heat transport with strong
dependence on the ratio of gaps and diffusion constants in different
bands.

%\textbf{Acknowledgements}
\begin{acknowledgments}
The authors would like to acknowledge useful discussions with
N.B.Kopnin, V.M.Vinokur, and I.I.Mazin. A.E.K. is supported by
UChicago Argonne, LLC, operator of Argonne National Laboratory, a
U.S. Department of Energy Office of Science laboratory, operated
under contract No. DE-AC02-06CH11357. This work also was supported
by the \textquotedblleft Center for Emergent
Superconductivity\textquotedblright, an Energy Frontier Research
Center funded by the U.S. Department of Energy, Office of Science,
Office of Basic Energy Sciences under Award Number DE-AC0298CH1088.
\end{acknowledgments}

\end{document}